# *Ab initio* study of electron-phonon interaction in phosphorene

*Bolin Liao[1], Jiawei Zhou[1], Bo Qiu[1], Mildred S. Dresselhaus[2,3*] and Gang Chen[1*]*

1. Department of Mechanical Engineering, Massachusetts Institute of Technology, Cambridge, Massachusetts, 02139, USA

2. Department of Electrical Engineering and Computer Science, Massachusetts Institute of Technology, Cambridge, Massachusetts, 02139, USA

3. Department of Physics, Massachusetts Institute of Technology, Cambridge, Massachusetts, 02139, USA

**Abstract**

The monolayer of black phosphorous, or "phosphorene", has recently emerged as a new 2D semiconductor with intriguing highly anisotropic transport properties. Existing calculations of its intrinsic phonon-limited electronic transport properties so far rely on the deformation potential approximation, which is in general not directly applicable to anisotropic materials since the deformation along one specific direction can scatter electrons traveling in all directions. We perform a first-principles calculation of the electron-phonon interaction in phosphorene based on density functional perturbation theory and Wannier interpolation. Our calculation reveals that 1) the high anisotropy provides extra phase space for electron-phonon scattering, and 2) optical phonons have appreciable contributions. Both effects cannot be captured by the deformation potential calculations. Our simulation predicts carrier mobilities ~170 cm$^2$/Vs for both electrons and holes at 300K, and a thermoelectric figure of merit zT of up to 0.31 in p-type impurity-free phosphorene at 500K.

**Introduction**

20 years ago, Hicks and Dresselhaus predicted that low dimensional conductors could have better thermoelectric performance compared with their 3-dimensional bulk counterparts, mainly owing to the quantum confinement effect[1,2]. In particular, the electronic density of states in low-dimensional systems usually exhibit sharp changes with respect to the carrier energy, which is of significant benefit for improving the Seebeck coefficient[3]. In the past two decades, researchers have widely utilized the much-advanced nanotechnology to boost the thermoelectric performance using various approaches[4,5]. Experimentally, artificial low-dimensional structures, such as quantum dots[6], quantum wells[7], superlattices[8,9], a 2-dimensional electron gas[10] and nanowires[11,12] have all been studied for enhanced thermoelectric properties. These structures, however, are difficult to scale up due to the complexity of the fabrication process and the high cost.

Immediately after the first successful isolation of graphene[13], a stable monolayer of carbon, the look into natural low-dimensional conductors for good thermoelectrics started, in the hope that these materials are easier to obtain at a lower cost[14–16]. Unfortunately, graphene itself turns out to be a not-so-good thermoelectric material, for 1) the high electron-hole symmetry and the absence of a bandgap, which results in large detrimental bipolar conduction; 2) the linear dispersion for low-energy excitations, which leads to a smooth quadratic density of states without the preferred sharp features; and 3) ultra-high lattice thermal conductivity[17,18], in part coming from the large contribution of the less-scattered flexural phonon mode[19] (there have been theoretical works suggesting that the classical size effect in nanostructured graphene can largely reduce its lattice thermal conductivity[20,21]). Along these lines, natural 2-dimensional materials with a sizable bandgap, quadratic low-energy dispersion, and suppressed flexural phonon modes have been sought as better candidates for thermoelectrics, in addition to the generally preferred high carrier mobility. Some subsequently synthesized/isolated 2-dimensional materials usually fit some of the above criteria, but for most cases also possess serious drawbacks. For example, a monolayer of transition-metal dichalcogenide $MoS_2$ comes with a bandgap while being limited by its relatively low carrier mobility[22,23]; silicene and germanene, monolayers of silicon and germanium atoms arranged in honeycomb lattices, possess similar low-energy electronic structures as that of graphene, with very small band gaps (a few meV), which only arise from spin-orbit coupling[24,25].

Recently, a new member of the 2-dimensional-material family, single layers of black phosphorus dubbed "phosphorene", has emerged and attracted intense research interest[26–34]. In a phosphorene layer, phosphorus atoms are arranged in a puckered honeycomb lattice[35] with low symmetry and high anisotropy. This hinge-like puckered structure leads to intriguing mechanical properties, such as a negative Poisson ratio[36]. The resulting electronic structure is also highly anisotropic, with a fundamental bandgap of 2 eV[37] that can be potentially tuned either by changing number of layers[26], controlling the edge termination and the width of a ribbon[38] or imposing a strain[39,40]. The low-energy dispersion is quadratic with very different effective masses along armchair and zigzag directions[37] for both electrons and holes. This anisotropic electronic structure is useful for thermoelectric materials, since in the direction with a smaller effective mass, the carrier mobility and thus the electrical conductivity can be high, while the larger effective mass along the other direction contributes to an overall large density of states that improves the Seebeck coefficient. Moreover, few-layer black phosphorous has been experimentally found to exhibit high carrier mobility, especially for holes[26–28,41,42], while theoretical calculations on single-layer phosphorene have suggested even higher values[37,43], and the possible tunability via applying a strain[44,45]. These features have stimulated lots of research efforts in evaluating the potential thermoelectric performance of phosphorene. Lv et al. calculated the thermoelectric power factor of phosphorene[46], and further showed that strain helps improve the thermoelectric performance of phosphorene via inducing band convergence[47]. Qin et al. simulated the lattice thermal conductivity of phosphorene from first-principles[48], showed that the thermal transport was also highly anisotropic, and revealed that the much reduced lattice thermal conductivity compared with graphene could be largely attributed to the suppressed flexural mode. Fei et al. pointed out that the directions with higher electrical and thermal conductivity, respectively, in phosphorene are orthogonal to each other, which leads to a promising thermoelectric figure of merit along the armchair direction, exceeding 2 at 500K[37].

Although aforementioned works have studied the thermoelectric properties of phosphorene in some detail, the treatment of electron-phonon interaction in phosphorene has been limited to the constant relaxation time approximation for calculating Seebeck coefficient, and the deformation potential approximation for calculating the electrical conductivity. In particular, the existing deformation potential calculations[37,43] obtained separate deformation potentials for different transport directions by deforming the lattice along that direction. The validity of such an approach is questionable because the deformation along one direction will scatter electrons going in all directions. In this paper, we study the electron-phonon interaction in phosphorene fully from first-principles. We find that the deformation potential calculations tend to overestimate the carrier mobility, and we fully assess the potential of phosphorene as a thermoelectric material based on our simulation results.

**Methods**

We first carry out the standard density functional theory calculation of the electronic structure of phosphorene after obtaining a fully relaxed crystal structure using the Quantum Espresso package[49] and a norm-conserving pseudopotential with the Perdew-Wang exchange-correlation functional within the local density approximation[50]. We use $200 \times 200$ k-mesh and the triangular integration method[51] to generate an accurate electronic density of states. The phonon dispersion and the electron-phonon scattering matrix elements are calculated within density functional perturbation theory[52], initially on a coarse $10 \times 10$ q-mesh, and then along with the electronic structure on a coarse $10 \times 10$ k-mesh, are interpolated using the EPW package[53,54] to a dense $300 \times 300$ k-mesh covering half of the Brillouin zone centered around the $\Gamma$ point and $300 \times 300$ q-mesh in the full Brillouin zone using maximally localized Wannier functions[55] for calculating the electronic relaxation time due to electron-phonon interaction, which is given by the Fermi's golden rule[56] as

$$\frac{1}{\tau_{ep}(\mathbf{k})} = \frac{2\pi}{\hbar} \sum_{\mathbf{k'}} \left| \langle \mathbf{k'} | \partial_q V | \mathbf{k} \rangle \right|^2 \left[ (f_{\mathbf{k'}} + n_{\mathbf{q}}) \delta(E_{\mathbf{k}} - E_{\mathbf{k'}} + \hbar\omega_{\mathbf{q}}) \delta_{\mathbf{k+q,k'+G}} + (1 + n_{\mathbf{q}} - f_{\mathbf{k'}}) \delta(E_{\mathbf{k}} - E_{\mathbf{k'}} - \hbar\omega_{\mathbf{q}}) \delta_{\mathbf{k-q,k'+G}} \right] \left( 1 - \frac{\mathbf{v}_{\mathbf{k'}} \cdot \mathbf{v}_{\mathbf{k}}}{|\mathbf{v}_{\mathbf{k'}}||\mathbf{v}_{\mathbf{k}}|} \right), \quad (1)$$

where $\mathbf{k}$, $\mathbf{k'}$ and $\mathbf{q}$ are wavevectors of the initial and final electronic states and the participating phonon state, $E_{\mathbf{k}}$, $E_{\mathbf{k'}}$ and $\hbar\omega_{\mathbf{q}}$ are their energies, $f_{\mathbf{k}}$, $f_{\mathbf{k'}}$ and $n_{\mathbf{q}}$ are their equilibrium distribution functions, $\langle \mathbf{k'} | \partial_q V | \mathbf{k} \rangle$ is the electron-phonon matrix element, $\mathbf{G}$ is a reciprocal lattice vector, and $\mathbf{v}_{\mathbf{k}}$ and $\mathbf{v}_{\mathbf{k'}}$ are the group velocities of the initial and the final states. The factor $\left( 1 - \frac{\mathbf{v}_{\mathbf{k'}} \cdot \mathbf{v}_{\mathbf{k}}}{|\mathbf{v}_{\mathbf{k'}}||\mathbf{v}_{\mathbf{k}}|} \right)$ takes into account the momentum loss in the scattering processes, and the thus defined relaxation time is usually named "the momentum relaxation time", and used for calculating transport properties[57,58]. The summation in Eq. (1) is performed using the triangular method[51] to improve the convergence, and eliminate the need of choosing the Gaussian broadening parameter when doing the summation using Gaussian functions to approximate the delta functions. The calculated electronic relaxation times are then plugged into the standard formulae of the transport properties based on the Boltzmann transport equation[59]. The whole process is parameter-free and has been applied to study the thermoelectric transport properties of silicon by the authors with remarkably good agreement with experiments[60]. A similar calculation scheme has been applied by other researchers to studying the electron-phonon interactions in graphene[61,62].

## Results and Discussions

For brevity, the lattice parameters, the electronic structure and phonon dispersion are not shown here, since our results are essentially the same as previous reports. The electronic band gap is underestimated to be 0.8 eV, and is a well-known problem of density functional theory. The more accurate band gap of 2 eV from GW calculation[37] is imposed in the following calculations by rigidly shifting the bands. Figure 1 shows the electronic density of states, where well-defined step-like features specific to 2-dimensional quadratic bands are observed, as well as quasi-1-dimensional peaks near the band edges as a result of the high anisotropy, since both the lowest conduction band and the highest valence band are very flat along the zigzag direction, resembling 1-dimensional bands along the armchair direction, which signals a high Seebeck coefficient. This feature is reminiscent of the quasi-2-dimensional bands in good bulk thermoelectrics, such as PbTe and PbSe[63].

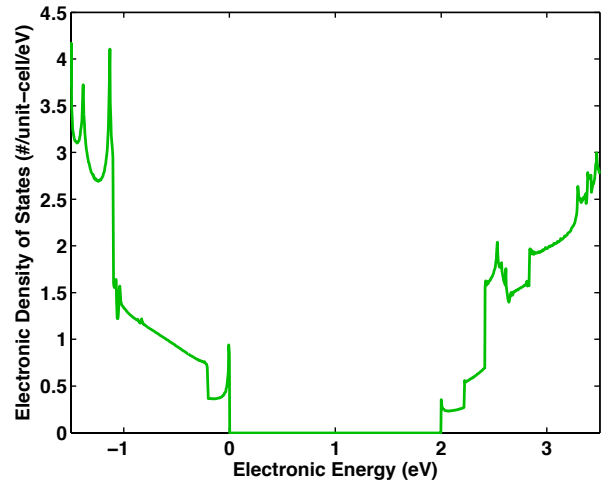

**Figure 1.** The electronic density of states of phosphorene. Step-like features characteristic to 2-dimensional quadratic bands are present, as well as quasi-1-dimensional peaks, which reflect the fact that both the conduction band and valence band are very flat along the zigzag direction, resembling 1-dimensional electronic bands.

We proceed to calculate the scattering rates and mobility for free carriers in phosphorene limited by electron-phonon interactions. We exclude contributions from the flexural phonon modes for two reasons: 1) out-of-plane vibrations do not contribute to the first-order electron-phonon interactions in 2D materials due to the inversion symmetry with respect to the material plane, as in the case of graphene[62,64] (although the flexural phonon modes in phosphorene have small in-plane components as well, their contributions are negligibly small); 2) so far in most experiments phosphorene samples are studied on a substrate, by which the flexural phonons will be largely suppressed. The scattering rates are presented in Fig. 2, and compared with the deformation potential calculation using parameters from Qiao et al.[43]. Although the deformation potential approximation is not rigorously applicable in this case due to previously mentioned reasons, in general it gives reasonable estimates of the average strength of electron-acoustic-phonon interactions, except for the case of holes in the zigzag direction, where the predicted scattering rate using the deformation potential approximation is 3 orders of magnitude lower than our result, and a hole mobility of 26,000 $cm^2$/Vs was predicted accordingly[43]. Our results indicate that the contributions from optical phonons are not negligible, especially for carriers slightly away from the band edges. More importantly, we observe peak structures of the scattering rates near the band edges, similar to the density of states. This can be explained by the high anisotropy of the band structure. According to Eq. (1), the scattering rate of a specific carrier state depends on the available phase space for the final states. In other words, the large number of carrier states along the zigzag direction provides a large number of available final states for carriers traveling along the armchair direction to be scattered to. In this way the carrier transport along the two directions are coupled through the electron-phonon interaction and the electron-phonon scattering rates follow the trend of the total density of states.

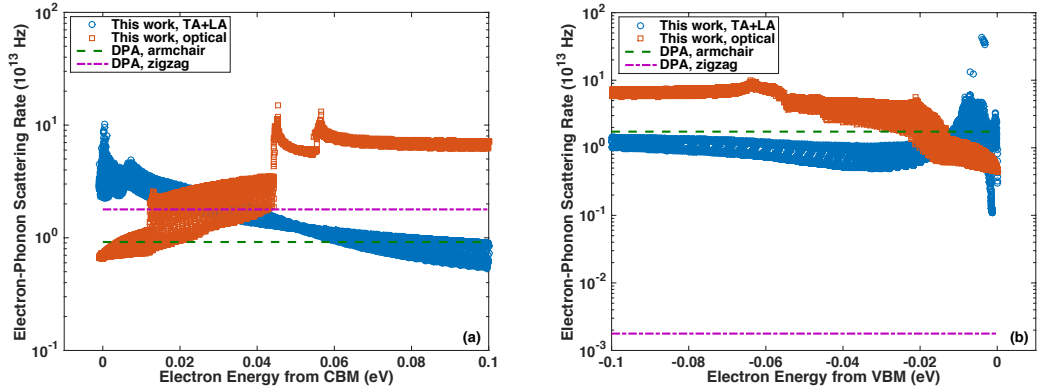

**Figure 2.** The calculated electron-phonon scattering rates (circles for acoustic phonon scattering and squares for optical phonon scattering) for (a) electrons and (b) holes, compared to the deformation potential approximation (DPA) results (dash and dash-dot lines).

The peak structures of the scattering rates near the band edges are expected to have a major impact on the carrier mobility since carriers near the band edges contribute the most to the transport. We show here in Fig. 3(a) the carrier mobilities of electrons and holes along armchair directions at 300K, with respect to the carrier concentration. We simulate the effect of carrier concentration by rigidly shifting the Fermi level, assuming the electronic band structure is not greatly affected by free carriers. We predict the phonon-limited carrier mobility of phosphorene is ~170 cm$^2$/Vs for both electrons and holes at 300K. Experimentally Xia et al.[28] measured the hole mobility of a 15 nm thick (~30 atomic layers) black phosphorous sample to be ~600 cm$^2$/Vs and that of an 8 nm thick (~15 atomic layers) sample to be ~400 cm$^2$/Vs, and more recently Xiang et al.[42] measured the hole mobility of a 4.8 nm thick (~8 atomic layers) sample to be ~200 cm$^2$/Vs. The decreasing trend of the hole mobility with decreasing number of atomic layers was previously attributed to the sample quality degradation[28]. Our finding suggests that the drastically increased anisotropy with decreasing number of atomic layers[43] could also contribute to the observed reduction of the carrier mobility. In Fig. 3(b) we present the calculated electronic thermal conductivity for p-type phosphorene along the armchair direction, which becomes appreciable at higher doping concentrations. More detailed information and data including transport data along the zigzag direction can be found in the Supplementary Material.

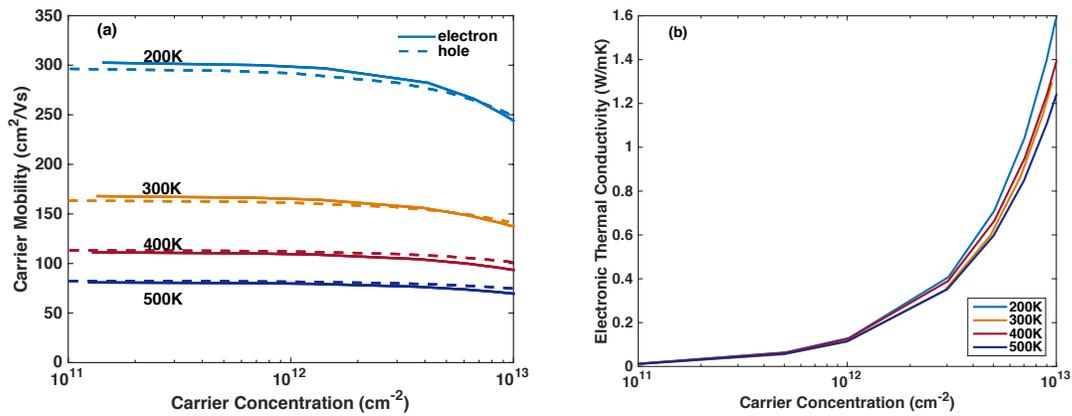

**Figure 3.** (a) The calculated carrier mobility for electrons (solid line) and holes (dashed line) along the armchair direction at different temperatures; (b) the calculated electronic thermal conductivity along the armchair direction for p-type phosphorene at different temperatures; all plotted versus the carrier concentration.

In Fig. 4 we show the calculated Seebeck coefficient and thermoelectric power factor along the armchair direction for holes (since the p-type phosphorene has better thermoelectric performance, we put all data for n-type phosphorene in the Supplementary Material), at various carrier concentrations and temperatures. In calculating the electrical conductivity, the thickness of the phosphorene sheet is chosen as the interlayer distance in bulk phosphorous, 0.55 nm[37]. Although this conventional choice seems somewhat arbitrary, it will not affect the thermoelectric figure of merit zT because the same factor appears in the thermal conductivity as well. Owing to the special features of the electronic structure mentioned above, the Seebeck coefficient is high, and the thermoelectric power factor reaches ~70 µW/cm-K$^2$ in p-type phosphorene at room temperature. This number is comparable to that in state-of-the-art bulk thermoelectric materials, such as BiTeSb alloy[65] (the arbitrariness of choosing the film thickness may render this comparison unfair, to some extent).

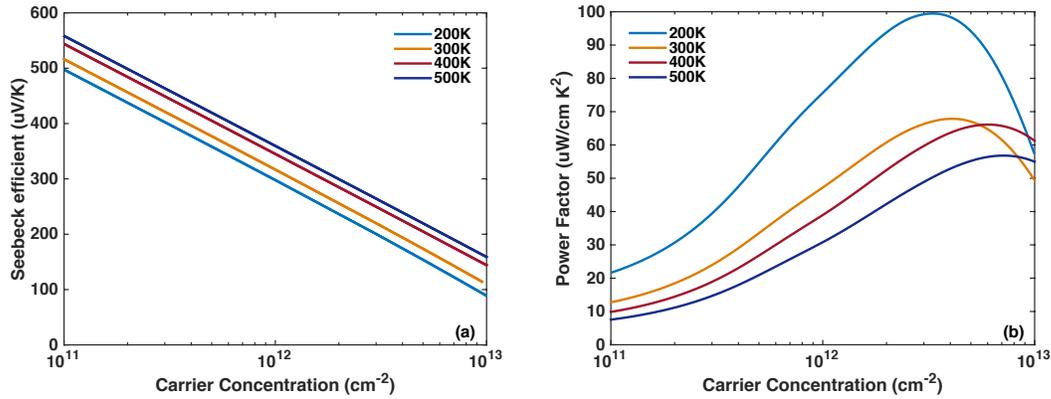

**Figure 4.** The calculated (a) Seebeck coefficient and (b) thermoelectric power factor versus the carrier concentration for holes along the armchair direction at different temperatures.

With the recently reported phonon thermal conductivity of phosphorene calculated from first-principles[48], we have all the ingredients for calculating the thermoelectric figure of merit zT, and the results are shown in Fig. 5 for p-type phosphorene along the armchair direction at temperatures up to 500K. The optimal zT is ~0.15 at 300K and ~0.31 at 500K for p-type, with the optimal carrier concentration around $3\times10^{12}$ cm$^{-2}$ at 300K and $8\times10^{12}$ cm$^{-2}$ at 500K. These values are for impurity-free phosphorene and should be regarded as an upper limit for the thermoelectric performance of phosphorene. To evaluate the potential effectiveness of the nanostructuring approach[65] for further improving zT, we calculate the accumulated contribution to the transport properties along the armchair direction from individual carrier states with respect to their mean free paths, as shown in Fig. 6, where we choose the optimal carrier concentration $3\times10^{12}$ cm$^{-2}$ at 300K. The calculated accumulated contribution represents fictitious values of the transport properties if carriers with mean free paths longer than a certain value are strongly suppressed by nanostructures (removed from the calculation), which is then normalized by the corresponding bulk values. It estimates the relative effectiveness of nanostructures with a characteristic size in affecting the transport properties. Figure 6 indicates that major contribution to the transport comes from carriers with mean free paths below 10 nm at 300K. Since the phonon thermal conductivity has contributions from phonons with mean free paths up to 1 μm[48], nanostructures with a characteristic size of ~10 nm can significantly reduce the phonon thermal conductivity and preserve the electronic properties. In this ideal case the figure of merit zT at 300K can be improved to ~0.75 in p-type phosphorene (in reality phonons with mean free paths longer than 10nm still conduct some heat, just with shorter mean free paths).

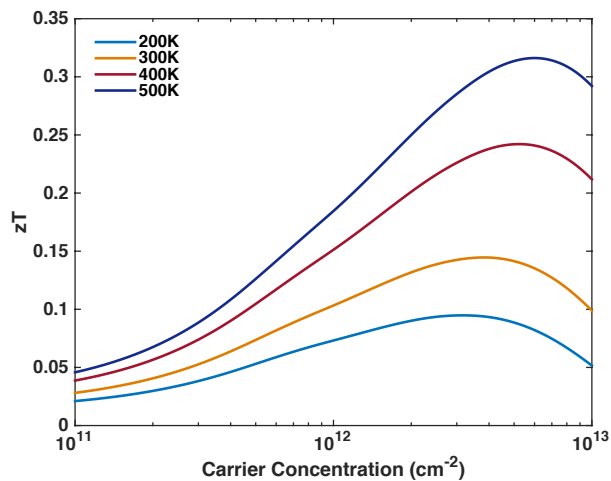

**Figure 5.** Thermoelectric figure of merit zT versus the carrier concentration for p-type phosphorene along the armchair direction at different temperatures, limited by the electron-phonon scattering.

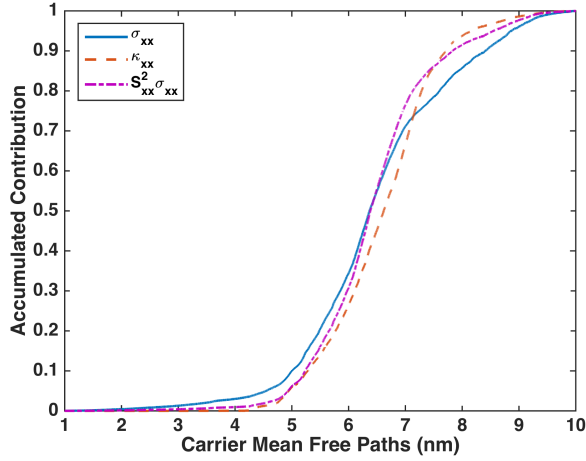

**Figure 6.** Accumulated contribution to transport properties ($\sigma_{xx}$: electrical conductivity, $S_{xx}$: Seebeck coefficient, $\kappa_{xx}$: electronic thermal conductivity) along the armchair direction from individual carrier states with respect to their mean free paths in p-type phosphorene. The carrier concentration is at $3\times 10^{12}$ cm$^{-2}$ and the temperature is at 300K.

**Conclusion**

In summary, we study the potential thermoelectric performance of phosphorene via first-principles calculation of the electron-phonon interactions. Our calculation finds that previous deformation potential calculations overestimate the carrier mobility due to high anisotropy and contributions from optical phonons. We further calculate the figure of merit zT for phosphorene and the carrier mean free path distribution, which, in comparison to the phonon mean free path distribution, indicates that nanostructuring can effectively enhance the thermoelectric performance of phosphorene.

ASSOCIATED CONTENT

**Supporting Information Available:** Calculation results for transport properties along the zigzag direction and in n-type phosphorene; formulae for calculating transport properties. This material is available free of charge via the Internet at http://pubs.acs.org.

AUTHOR INFORMATION

**Corresponding Authors**

*Email: millie@mgm.mit.edu (M.S.D)., *Email: gchen2@mit.edu (G.C.).

**Author Contributions**

The manuscript was written through contributions of all authors. All authors have given approval to the final version of the manuscript.

ACKNOWLEDGMENT

We thank Sangyeop Lee, Xi Ling, Albert Liao and Cheol-Hwan Park for helpful discussions. This article is based upon work supported partially by S$^3$TEC, an Energy Frontier Research Center funded by the U.S. Department of Energy, Office of Basic Energy Sciences, under Award No. DE-FG02-09ER46577 (for potential thermoelectric power generation using phosphorene), and partially by the Air Force Office of Scientific Research Multidisciplinary Research Program of the University Research Initiative

(AFOSR MURI) via Ohio State University, Contract FA9550-10-1-0533 (for potential thermoelectric cooling using phosphorene).**REFERENCES**

(1)  Hicks, L. D.; Dresselhaus, M. S. *Phys. Rev. B* **1993**, *47*, 12727–12731.

(2)  Hicks, L. D.; Dresselhaus, M. S. *Phys. Rev. B* **1993**, *47*, 16631–16634.

(3)  Mahan, G. D.; Sofo, J. O. *Proc. Natl. Acad. Sci.* **1996**, *93*, 7436–7439.

(4)  Dresselhaus, M. S.; Chen, G.; Tang, M. Y.; Yang, R. G.; Lee, H.; Wang, D. Z.; Ren, Z. F.; Fleurial, J.-P.; Gogna, P. *Adv. Mater.* **2007**, *19*, 1043–1053.

(5)  Zebarjadi, M.; Esfarjani, K.; Dresselhaus, M. S.; Ren, Z. F.; Chen, G. *Energy Environ. Sci.* **2012**, *5*, 5147–5162.

(6)  Harman, T. C.; Taylor, P. J.; Walsh, M. P.; LaForge, B. E. *Science* **2002**, *297*, 2229–2232.

(7)  Hicks, L. D.; Harman, T. C.; Sun, X.; Dresselhaus, M. S. *Phys. Rev. B* **1996**, *53*, R10493–R10496.

(8)  Venkatasubramanian, R.; Siivola, E.; Colpitts, T.; O'Quinn, B. *Nature* **2001**, *413*, 597–602.

(9)  Chowdhury, I.; Prasher, R.; Lofgreen, K.; Chrysler, G.; Narasimhan, S.; Mahajan, R.; Koester, D.; Alley, R.; Venkatasubramanian, R. *Nat. Nanotechnol.* **2009**, *4*, 235–238.

(10) Ohta, H.; Kim, S.; Mune, Y.; Mizoguchi, T.; Nomura, K.; Ohta, S.; Nomura, T.; Nakanishi, Y.; Ikuhara, Y.; Hirano, M.; Hosono, H.; Koumoto, K. *Nat. Mater.* **2007**, *6*, 129–134.

(11) Boukai, A. I.; Bunimovich, Y.; Tahir-Kheli, J.; Yu, J.-K.; Goddard Iii, W. A.; Heath, J. R. *Nature* **2008**, *451*, 168–171.

(12) Hochbaum, A. I.; Chen, R.; Delgado, R. D.; Liang, W.; Garnett, E. C.; Najarian, M.; Majumdar, A.; Yang, P. *Nature* **2008**, *451*, 163–167.

(13) Novoselov, K. S.; Geim, A. K.; Morozov, S. V.; Jiang, D.; Zhang, Y.; Dubonos, S. V.; Grigorieva, I. V.; Firsov, A. A. *Science* **2004**, *306*, 666–669.

(14) Zuev, Y. M.; Chang, W.; Kim, P. *Phys. Rev. Lett.* **2009**, *102*, 096807.

(15) Wei, P.; Bao, W.; Pu, Y.; Lau, C. N.; Shi, J. *Phys. Rev. Lett.* **2009**, *102*, 166808.

(16) Wang, C.-R.; Lu, W.-S.; Hao, L.; Lee, W.-L.; Lee, T.-K.; Lin, F.; Cheng, I.-C.; Chen, J.-Z. *Phys. Rev. Lett.* **2011**, *107*, 186602.

(17) Balandin, A. A.; Ghosh, S.; Bao, W.; Calizo, I.; Teweldebrhan, D.; Miao, F.; Lau, C. N. *Nano Lett.* **2008**, *8*, 902–907.

(18) Seol, J. H.; Jo, I.; Moore, A. L.; Lindsay, L.; Aitken, Z. H.; Pettes, M. T.; Li, X.; Yao, Z.; Huang, R.; Broido, D.; Mingo, N.; Ruoff, R. S.; Shi, L. *Science* **2010**, *328*, 213–216.

(19) Lindsay, L.; Broido, D. A.; Mingo, N. *Phys. Rev. B* **2010**, *82*, 115427.

(20) Kim, J. Y.; Lee, J.-H.; Grossman, J. C. *ACS Nano* **2012**, *6*, 9050–9057.

(21) Sevinçli, H.; Sevik, C.; Çağın, T.; Cuniberti, G. *Sci. Rep.* **2013**, *3*, 1228.

(22) Radisavljevic, B.; Radenovic, A.; Brivio, J.; Giacometti, V.; Kis, A. *Nat. Nanotechnol.* **2011**, *6*, 147–150.

(23) Yoon, Y.; Ganapathi, K.; Salahuddin, S. *Nano Lett.* **2011**, *11*, 3768–3773.

(24) Cahangirov, S.; Topsakal, M.; Aktürk, E.; Şahin, H.; Ciraci, S. *Phys. Rev. Lett.* **2009**, *102*, 236804.

(25) Tsai, W.-F.; Huang, C.-Y.; Chang, T.-R.; Lin, H.; Jeng, H.-T.; Bansil, A. *Nat. Commun.* **2013**, *4*, 1500.

(26) Liu, H.; Neal, A. T.; Zhu, Z.; Luo, Z.; Xu, X.; Tománek, D.; Ye, P. D. *ACS Nano* **2014**, *8*, 4033–4041.

(27) Li, L.; Yu, Y.; Ye, G. J.; Ge, Q.; Ou, X.; Wu, H.; Feng, D.; Chen, X. H.; Zhang, Y. *Nat. Nanotechnol.* **2014**, *9*, 372–377.

(28) Xia, F.; Wang, H.; Jia, Y. *Nat. Commun.* **2014**, *5*, 4458.

(29) Low, T.; Engel, M.; Steiner, M.; Avouris, P. *Phys. Rev. B* **2014**, *90*, 081408.

(30) Engel, M.; Steiner, M.; Avouris, P. *Nano Lett.* **2014**, *14*, 6414–6417.

(31) Wang, H.; Wang, X.; Xia, F.; Wang, L.; Jiang, H.; Xia, Q.; Chin, M. L.; Dubey, M.; Han, S. *Nano Lett.* **2014**, *14*, 6424–6429.

(32) Ziletti, A.; Carvalho, A.; Campbell, D. K.; Coker, D. F.; Castro Neto, A. H. *Phys. Rev. Lett.* **2015**, *114*, 046801.

(33) Ling, X.; Liang, L.; Huang, S.; Puretzky, A. A.; Geohegan, D. B.; Sumpter, B. G.; Kong, J.; Meunier, V.; Dresselhaus, M. S. *arXiv:1502.07804* **2015**.

(34) Liu, Q.; Zhang, X.; Abdalla, L. B.; Fazzio, A.; Zunger, A. *Nano Lett*. **2015**, *15*, 1222.

(35) Asahina, H.; Shindo, K.; Morita, A. *J. Phys. Soc. Jpn*. **1982**, *51*, 1193–1199.

(36) Jiang, J.-W.; Park, H. S. *Nat. Commun*. **2014**, *5*, 4727.

(37) Fei, R.; Faghaninia, A.; Soklaski, R.; Yan, J.-A.; Lo, C.; Yang, L. *Nano Lett*. **2014**, *14*, 6393–6399.

(38) Ramasubramaniam, A.; Muniz, A. R. *Phys. Rev. B* **2014**, *90*, 085424.

(39) Çakır, D.; Sahin, H.; Peeters, F. M. *Phys. Rev. B* **2014**, *90*, 205421.

(40) Elahi, M.; Khaliji, K.; Tabatabaei, S. M.; Pourfath, M.; Asgari, R. *Phys. Rev. B* **2015**, *91*, 115412.

(41) Koenig, S. P.; Doganov, R. A.; Schmidt, H.; Neto, A. H. C.; Özyilmaz, B. *Appl. Phys. Lett*. **2014**, *104*, 103106.

(42) Xiang, D.; Han, C.; Wu, J.; Zhong, S.; Liu, Y.; Lin, J.; Zhang, X.-A.; Ping Hu, W.; Özyilmaz, B.; Neto, A. H. C.; Wee, A. T. S.; Chen, W. *Nat. Commun*. **2015**, *6*, 6485.

(43) Qiao, J.; Kong, X.; Hu, Z.-X.; Yang, F.; Ji, W. *Nat. Commun*. **2014**, *5*, 4475.

(44) Fei, R.; Yang, L. *Nano Lett*. **2014**, *14*, 2884–2889.

(45) Morgan Stewart, H.; Shevlin, S. A.; Catlow, C. R. A.; Guo, Z. X. *Nano Lett.* **2015**, *15*, 2006–2010.

(46) Lv, H. Y.; Lu, W. J.; Shao, D. F.; Sun, Y. P. *arXiv:1404.5171* **2014**.

(47) Lv, H. Y.; Lu, W. J.; Shao, D. F.; Sun, Y. P. *Phys. Rev. B* **2014**, *90*, 085433.

(48) Qin, G.; Yan, Q.-B.; Qin, Z.; Yue, S.-Y.; Hu, M.; Su, G. *Phys. Chem. Chem. Phys.* **2015**, *17*, 4854–4858.

(49) Giannozzi, P. et al. *J. Phys. Condens. Matter* **2009**, *21*, 395502.

(50) Perdew, J. P.; Wang, Y. *Phys. Rev. B* **1992**, *45*, 13244–13249.

(51) Ashraff, J. A.; Loly, P. D. *J. Phys. C Solid State Phys.* **1987**, *20*, 4823.

(52) Baroni, S.; de Gironcoli, S.; Dal Corso, A.; Giannozzi, P. *Rev. Mod. Phys.* **2001**, *73*, 515–562.

(53) Giustino, F.; Cohen, M. L.; Louie, S. G. *Phys. Rev. B* **2007**, *76*, 165108.

(54) Noffsinger, J.; Giustino, F.; Malone, B. D.; Park, C.-H.; Louie, S. G.; Cohen, M. L. *Comput. Phys. Commun.* **2010**, *181*, 2140–2148.

(55) Marzari, N.; Mostofi, A. A.; Yates, J. R.; Souza, I.; Vanderbilt, D. *Rev. Mod. Phys.* **2012**, *84*, 1419–1475.

(56) Ziman, J. M. *Electrons and Phonons: The Theory of Transport Phenomena in Solids*; Clarendon Press: Oxford, 1960.

(57) Datta, S. *Electronic Transport in Mesoscopic Systems*; Cambridge University Press: Cambridge, 1997.

(58) Lundstrom, M. *Fundamentals of Carrier Transport*; Cambridge University Press: New York, 2009.

(59) Chen, G. *Nanoscale energy transport and conversion: a parallel treatment of electrons, molecules, phonons, and photons*; Oxford University Press: Oxford; New York, 2005.

(60) Qiu, B.; Tian, Z.; Vallabhaneni, A.; Liao, B.; Mendoza, J. M.; Restrepo, O. D.; Ruan, X.; Chen, G. *arXiv:1409.4862* **2014**.

(61) Park, C.-H.; Giustino, F.; Cohen, M. L.; Louie, S. G. *Phys. Rev. Lett.* **2007**, *99*, 086804.

(62) Park, C.-H.; Bonini, N.; Sohier, T.; Samsonidze, G.; Kozinsky, B.; Calandra, M.; Mauri, F.; Marzari, N. *Nano Lett.* **2014**, *14*, 1113–1119.

(63) Parker, D.; Chen, X.; Singh, D. J. *Phys. Rev. Lett.* **2013**, *110*, 146601.

(64) Castro, E. V.; Ochoa, H.; Katsnelson, M. I.; Gorbachev, R. V.; Elias, D. C.; Novoselov, K. S.; Geim, A. K.; Guinea, F. *Phys. Rev. Lett.* **2010**, *105*, 266601.

(65) Poudel, B.; Hao, Q.; Ma, Y.; Lan, Y.; Minnich, A.; Yu, B.; Yan, X.; Wang, D.; Muto, A.; Vashaee, D.; Chen, X.; Liu, J.; Dresselhaus, M. S.; Chen, G.; Ren, Z. *Science* **2008**, *320*, 634–638.

# Supplementary Material

**1. Carrier Mobility along the zigzag direction**

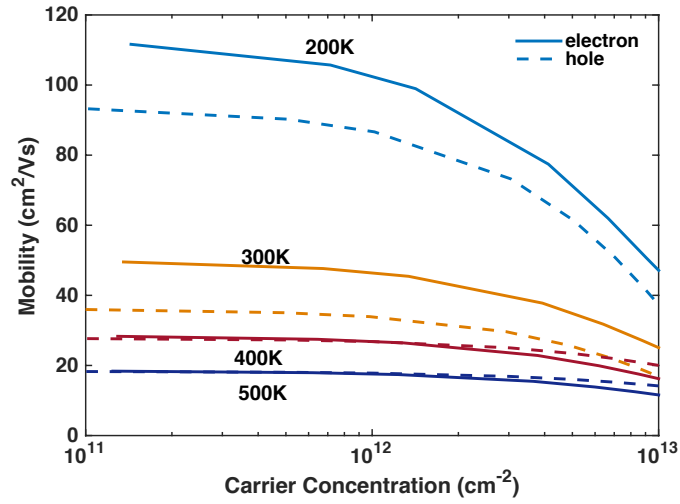

Figure S1. Calculated carrier mobility along the zigzag direction for electrons (solid lines) and holes (dashed lines).

## 2. Transport properties for n-type phosphorene

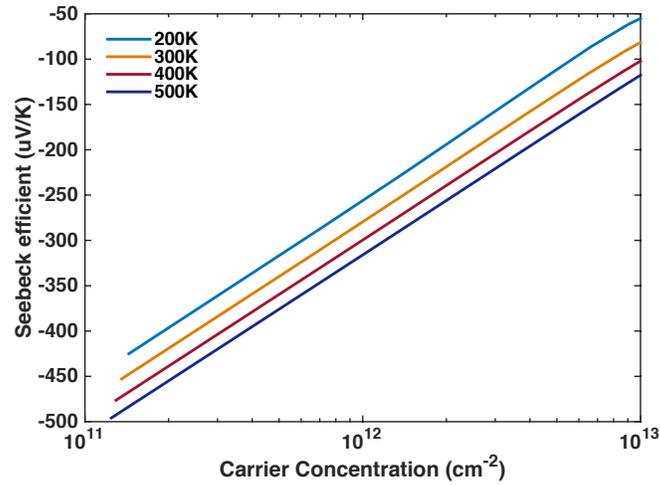

Figure S2 Calculated Seebeck coefficient along the armchair direction for n-type phosphorene versus the carrier concentration for various temperatures.

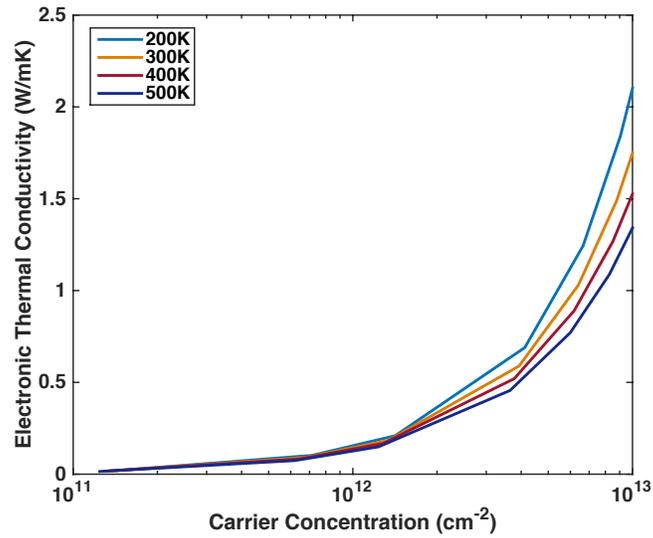

Figure S3 Calculated electronic thermal conductivity along the armchair direction for n-type phosphorene versus the carrier concentration for various temperatures.

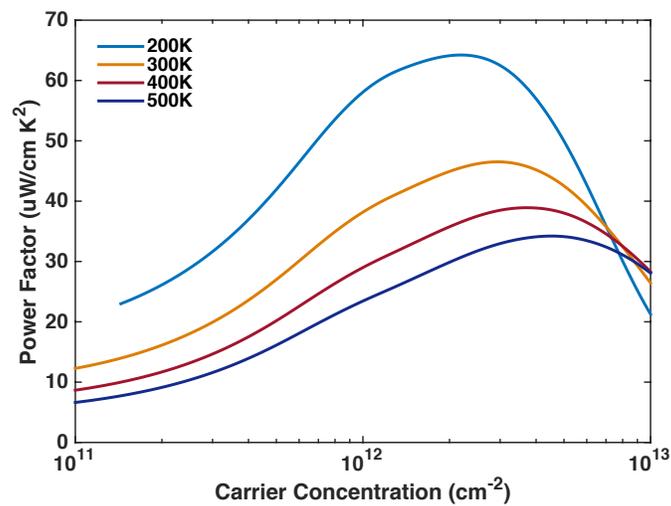

Figure S4 Calculated power factor along armchair direction for n-type phosphorene versus the carrier concentration for various temperatures.

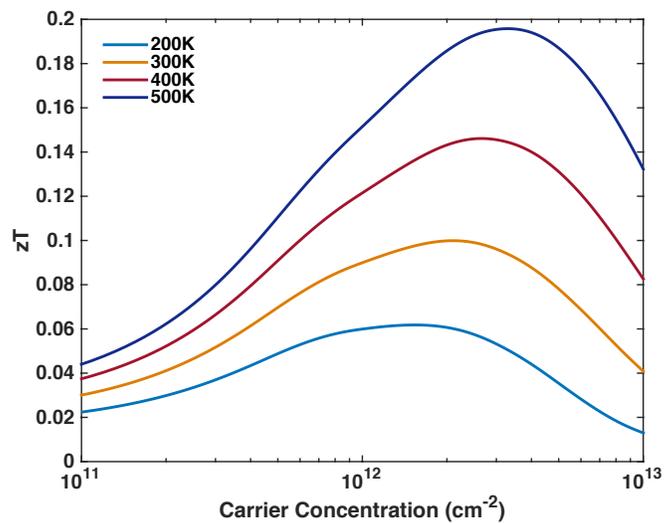

Figure S6. Calculated figure of merit zT along the armchair direction for n-type phosphorene versus the carrier concentration for various temperatures.

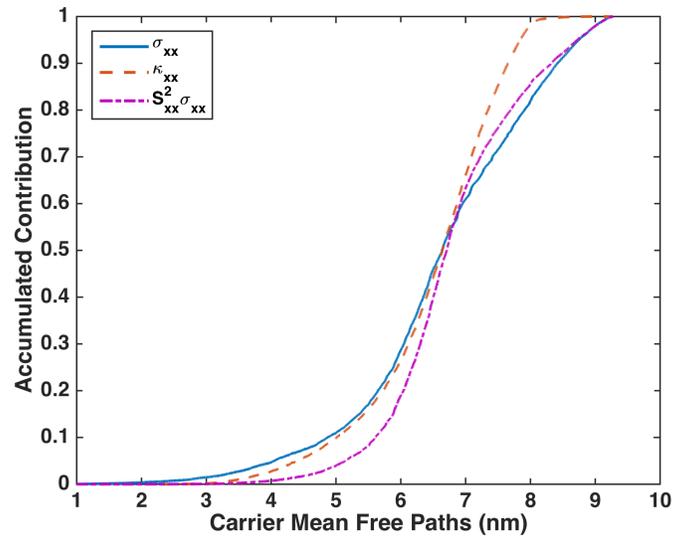

Figure S7. Calculated carrier mean free path distribution along the armchair direction for n-type phosphorene.

## 3. Formulae for transport properties

We use the following formulae for calculating transport properties:

$$\sigma_{xx} = \frac{1}{NS} \sum_{\mathbf{k}} -e^2 v_x^2 \tau_{ep}^{\mathbf{k}} \frac{\partial f_{\mathbf{k}}}{\partial E_{\mathbf{k}}}, \tag{S1}$$

$$S_{xx} = -\frac{1}{eT} \frac{\sum_{\mathbf{k}} (E_{\mathbf{k}} - E_f) v_x^2 \tau_{ep}^{\mathbf{k}} \frac{\partial f_{\mathbf{k}}}{\partial E_{\mathbf{k}}}}{\sum_{\mathbf{k}} v_x^2 \tau_{ep}^{\mathbf{k}} \frac{\partial f_{\mathbf{k}}}{\partial E_{\mathbf{k}}}}, \tag{S2}$$

$$\kappa_{xx} = \frac{1}{NS} \sum_{\mathbf{k}} -\frac{(E_{\mathbf{k}} - E_f)^2}{T} v_x^2 \tau_{ep}^{\mathbf{k}} \frac{\partial f_{\mathbf{k}}}{\partial E_{\mathbf{k}}} - TS_{xx}^2 \sigma_{xx}, \tag{S3}$$

where $\sigma_{xx}$ is the electrical conductivity, $S_{xx}$ is the Seebeck coefficient, $\kappa_{xx}$ is the zero-current electronic thermal conductivity, $N$ is the total number of k-points in the full Brillouin zone, $S$ is the area of a unit cell, $e$ is the electron charge, $v_x$ is the group velocity along x-direction, $\tau_{ep}^{\mathbf{k}}$ is the electron-phonon relaxation time, $f_{\mathbf{k}}$ is the Fermi-Dirac distribution, $E_{\mathbf{k}}$ is the electron energy and $T$ is temperature.